\documentclass[
  aps,
  prd,
  reprint,
  amsmath,
  amssymb,
  floatfix,
  longbibliography
]{revtex4-2}

\usepackage{amsthm,mathtools}
\usepackage{hyperref}
\hypersetup{colorlinks=true,linkcolor=blue,urlcolor=blue,citecolor=blue}
\usepackage{orcidlink}

\begin{document}

\title{Horizon Microstructure Thermodynamics in AdS Black Holes:\\
Smarr-Consistent Excitation Enthalpy}


\author{Juan Diego Haro\orcidlink{0009-0003-0932-7050}}
\affiliation{Departamento de F\'isica, Colegio de Ciencias e Ingenier\'ia, Universidad San Francisco de Quito,\\ Diego de Robles y V\'ia Interoce\'anica, Quito 17901, Ecuador}

\author{Ernesto Medina\orcidlink{0000-0002-1566-0170}}
\affiliation{Departamento de F\'isica, Colegio de Ciencias e Ingenier\'ia, Universidad San Francisco de Quito,\\ Diego de Robles y V\'ia Interoce\'anica, Quito 17901, Ecuador}
\affiliation{Laboratoire de Physique et Chimie Th\'eoriques, Universit\'e de Lorraine, CNRS, Nancy, France}
\affiliation{Arizona State University, School of Molecular Sciences, 551 E University Dr, Tempe, AZ 85281, USA}

\author{Bertrand Berche\orcidlink{0000-0002-4254-807X}}
\affiliation{Laboratoire de Physique et Chimie Th\'eoriques, Universit\'e de Lorraine, CNRS, Nancy, France}
\affiliation{Departamento de F\'isica, Colegio de Ciencias e Ingenier\'ia, Universidad San Francisco de Quito,\\ Diego de Robles y V\'ia Interoce\'anica, Quito 17901, Ecuador}

\author{Pedro Bargue\~no\orcidlink{0000-0001-5453-9042}}
\affiliation{Departamento de F\'isica, Universidad de Alicante, Campus de San Vicente del Raspeig, Alicante, E-03690, Spain}

\author{Ernesto Contreras\orcidlink{0000-0001-6403-4101}}
\affiliation{Departamento de F\'isica, Universidad de Alicante, Campus de San Vicente del Raspeig, Alicante, E-03690, Spain}

\begin{abstract}
In this work we formulate a horizon-microstructure description of four-dimensional AdS black holes in which
 a horizon of area $A$ is resolved into $\mathcal{N}=A/a_p$ microscopic sites and $N$ occupied horizon sites.
The central result is that the combinatorics of this partially occupied horizon sector yields the entropy directly:
in the finite-filling regime the leading term is proportional to the area, and the maximal-entropy filling reproduces the Bekenstein--Hawking law with $a_p=4\ln 2\  \ell_{p}^{2}$.
Subleading corrections include a subtractive logarithmic term and an inverse-area expansion.
We then show that this partially occupied regime admits a thermodynamic justification from an extended first law with chemical potential $\mu$, a Smarr-consistent excitation enthalpy $\delta M(A,P,N)$, and an AdS control parameter $u=PS$.
In this interpretation, the combinatorics provides the dominant horizon entropy, while the thermodynamic sector supplies a dressing that selects the equilibrium filling and assigns a finite excitation cost to departures from a reference partially occupied configuration.
\end{abstract}

\maketitle
\section{Introduction}

The connection between black hole (BH) horizon area and entropy, $S = k_B \frac{A}{4 \ell_p^2}$, is a cornerstone of the
unification of quantum theory and general relativity ~\cite{Bekenstein1973,Hawking1975} which usually serves to test against 
different models of quantum gravity. In addition, it has served as a key input to the complete derivation of Einstein equations from thermodynamic principles ~\cite{Jacobson1995} and to a insightful reformulation of classical gravity using thermodynamical
concepts (see, for example, ~\cite{Padmanabhan2010,Padmanabhan2010bis}). Beyond this doubly continuum description, where both spacetime and thermodynamical principles are 
deeply linked, the search for unambiguous degrees of freedom (dof) of BHs, together with their microscopical counting,
are still missing. While there have been considerable advances during the last years, mainly within string theory 
~\cite{STROMINGER1996}, loop quantum gravity (LQG) ~\cite{Kaul2000,Meissner2004}, AdS/CFT methods \cite{Carlip2000} and some models based on Wheeler's ``it from bit" idea ~\cite{Wheeler1989,Makela2019,Davidson2019,Bargueno2022,Bargueno2023}, there is no general consensus regarding the physical nature of the aforementioned dofs. However, what is generally accepted is that, when quantum corrections are switched on, BH entropy acquires a logarithmic correction ~\cite{Carlip2000,Kaul2000} {\it i.e.}, $S=k_{B}\left(\frac{A}{4 \ell_p^2 }+ c \; \ln \frac{A}{ \ell_p^2}\right)$, where $c$ is a model-dependent constant. Specifically, the values $c=-1/2$ and $-3/2$ are obtained from a $U(1)$ and $SU(2)$ formulation of LQG
(see ~\cite{Perez2017} for a review). Interestingly, the $c=-3/2$ was obtained by means of a CFT-based counting ~\cite{Carlip2000}. However, we have to point out that string theory calculations deviate from the previously mentioned values for $c$ ~\cite{Banerjee2011}. 
Very recently ~\cite{Bala2024,Bala2024bis} it has been shown that the Bekenstein-Hawking entropy emerges in a very general setting without any reference to AdS/CFT nor stringy hypothesis by assuming that (a) there is some UV completion, and (b) the  semiclassical Euclidean path integral provides sensible information about this completion. These results represent a significant step forward in showing that the Bekenstein-Hawking entropy can emerge from universal principles, without reference to a specific microscopic theory. However, their approach does not account for the subleading logarithmic correction. In this regard, we introduce a simple model that not only reproduces the Bekenstein-Hawking entropy, but also yields the logarithmic correction with coefficient $c = -1/2$, suggesting that this subleading term may also have a universal origin.
 As in \cite{Bala2024,Bala2024bis}, our model does not appeal to any specific microscopic realization of quantum gravity, reinforcing the idea that both the area law and its logarithmic correction may stem from universal features of discrete quantum geometry. 
 \\
 \\
 This work is organized as follows. In Sec.~\ref{sec:combi} we introduce the horizon microstructure model and derive the combinatorial entropy associated with partially occupied horizon sites. In Sec.~\ref{sec:thermo} we develop the thermodynamic framework, including the extended first law and the Smarr-consistent excitation enthalpy, and show how the partially occupied regime is thermodynamically justified. Sec.~\ref{sec:summ} summarizes the results and discuss their physical interpretation.

\section{Horizon microstructure: sites, occupation, and combinatorial entropy}
\label{sec:combi}

Let us start by introducing the essential assumptions of our model. We adopt a coarse-grained description of the horizon in which its total area $A$ is discretized into a set of fundamental patches of typical size $a_p\sim \ell_p^2$, where $\ell_p$ denotes the Planck length. Under this picture, the horizon can be regarded as a collection of independent microscopic sites, each contributing a quantum of area $a_p$. The total number of such sites is therefore proportional to the macroscopic area and is given by
\begin{equation}
\mathcal{N}(A)=\frac{A}{a_p}
\end{equation}
microscopic sites of area $a_p\sim \ell_p^2$.
We further assume that these sites can be either occupied or unoccupied according to some effective microscopic degrees of freedom living on the horizon, in analogy with a lattice gas model in the ideal solid solution approximation\cite{Bragg-Williams34,Bethe35,Peierls36,Nix-Shockley38}. Let $N$ denote the number of occupied horizon sites. The filling fraction is then naturally defined as the ratio between occupied and total sites,
\begin{equation}
n\equiv \frac{N}{\mathcal{N}}=\frac{N a_p}{A},
\qquad 0<n<1.
\end{equation}
Assuming that each site can be either occupied or empty, the microscopic degrees of freedom on the horizon can be modeled as a collection of two-state systems and under this assumption, the number of microstates at fixed $(A,N)$ is purely combinatorial and given by
\begin{equation}
\Omega(A,N)=\binom{\mathcal{N}}{N}.
\end{equation}
The associated microstructure entropy is then obtained from the standard Boltzmann relation ($k_{B}=1$)
\begin{equation}
S_{\rm micro}(A,N)=\ln\Omega(A,N)=\ln\binom{\mathcal{N}}{N}.
\end{equation}
A useful quantity for later thermodynamic considerations is the variation of the entropy with respect to the number of occupied sites. In the regime of large $\mathcal{N}$, where $N$ can be treated as a continuous variable, one finds
\begin{equation}
\left(\frac{\partial S_{\rm micro}}{\partial N}\right)_A
=\ln\!\left(\frac{\mathcal{N}-N}{N}\right)
=\ln\!\left(\frac{1-n}{n}\right).
\end{equation}


The main result of the construction is obtained directly from the combinatorics of occupied horizon sites.
We do not assume the Bekenstein--Hawking law as an input.
Instead, we ask what form the entropy takes when the occupied-site sector realizes a partially occupied regime with nontrivial finite filling,
\begin{equation}
n(A)\to n_0\in(0,1),
\qquad A\to\infty.
\end{equation}
In that regime, the Stirling expansion of the binomial coefficient gives the entropy without further thermodynamic assumptions. Specifically, for large $\mathcal{N}$ and $N=n\mathcal{N}$,
\begin{eqnarray}
\ln\binom{\mathcal{N}}{N}
&=&\mathcal{N}H(n)-\frac{1}{2}\ln\bigl(2\pi\mathcal{N}n(1-n)\bigr)\nonumber\\
&+&\frac{1}{12\mathcal{N}}\left(1-\frac{1}{n}-\frac{1}{1-n}\right)+O(\mathcal{N}^{-3}),
\end{eqnarray}
where
\begin{equation}
H(n)=-n\ln n-(1-n)\ln(1-n)
\end{equation}
is the binary entropy.
Evaluating at $n\to n_0$ and $\mathcal{N}=A/a_p$ gives
\begin{align}
S_{\rm micro}(A)
&=\frac{A}{a_p}H(n_0)-\frac{1}{2}\ln\!\left(\frac{A}{a_p}\right)
-\frac{1}{2}\ln\bigl(2\pi n_0(1-n_0)\bigr) \\
&\quad +\frac{a_p}{12A}\left[1-\frac{1}{n_0(1-n_0)}\right]+O(A^{-2}).
\end{align}
Thus the existence of a finite-filling regime already fixes the functional dependence of the entropy: an extensive area term, a subtractive logarithmic correction, and an inverse-area series. Even more, if we identify the leading extensive term with the Bekenstein--Hawking entropy, then
\begin{equation}
\frac{A}{a_p}H(n_0)=\frac{A}{4},
\end{equation}
which implies
\begin{equation}
a_p=4H(n_0).
\end{equation}
It is worth noting that the binary entropy reaches its maximum at
\begin{equation}
n_0=\frac{1}{2},
\qquad
H\!\left(\frac{1}{2}\right)=\ln 2,
\end{equation}
corresponding to a maximally disordered configuration \cite{Callen:450289} where occupied and empty sites are equally probable. In this case, the fundamental area scale becomes
\begin{equation}
a_p=(4\ln 2) \;\ell_p^2. \label{ap}
\end{equation}

At this point, some comments are in order. Bekenstein and Mukhanov argued \cite{BEKENSTEIN19957} that the black hole horizon area is an adiabatic invariant and should therefore be quantized as $A_n=\epsilon\, n\, \ell_p^2$. Using the Bekenstein--Hawking relation $S=\frac{A}{4\ell_p^2}$, this implies $S_n= \frac{\epsilon\; n}{4}$. Assuming that each level has a degeneracy $g_n=k^n$, with $k\in\mathbb{N}$, we get $S_n=\ln g_n=n\ln k$. Equating both expressions yields $\epsilon=4\ln k$, and hence an equally spaced area spectrum $A_n=4\,\ell_p^2(\ln k)\,n$. Even more, recent works based on the Weinstein integer for calculating black hole entropy \cite{Bargueno2023} arrived at the same conclusion regarding the minimum area. In this sense, Eq. (\ref{ap}) is in complete agreement with previous findings. In addition, the $-1/2 \ln$ correction appears when the classical degrees of freedom within LQG are described dynamically by a Chern-Simons theory in its $U(1)$ gauge \cite{Perez2017}.

With this identification, the entropy takes the form
\begin{eqnarray}
S(A)&=&\frac{A}{4\; \ell_p^2}-\frac{1}{2}\ln \frac{A}{a_p}-\frac{1}{2}\ln \bigl(2\pi n_0(1-n_0)\bigr)\nonumber\\
&&+\frac{D_1}{A}+O(A^{-3}),
\end{eqnarray}
where
\begin{equation}
D_1=\frac{a_p}{12}\left[1-\frac{1}{n_0(1-n_0)}\right].
\end{equation}
Equivalently,
\begin{equation}
S(A)=\frac{A}{4 \; \ell_p^2}-\frac{1}{2}\ln \frac{A}{a_p}+\sum\limits_{\alpha}\frac{D_{\alpha}}{A^{\alpha}}+\textnormal{constant},
\end{equation}
with higher Stirling terms generating $\alpha=1,3,5,\ldots$. For fixed filling $n_0$ the Stirling expansion (purely combinatorial) gives this signature inverse odd power series. More general approaches to the limiting filling regarding how $n(A)\rightarrow n_0$ will generate additional inverse-area corrections.

Summarizing, the factor $1/4$ in the Bekenstein--Hawking entropy emerges from the maximal value of the binary entropy $H(n)$ at half filling, $H(1/2)=\ln 2$, which fixes the elementary area scale, while the logarithmic correction arises from subleading terms in the Stirling expansion of the combinatorial entropy; however, the underlying dynamical mechanism that selects this maximal-entropy configuration remains unknown. In contrast with previous approaches where the filling is either fixed a priori, as in ``it from bit'' models \cite{Wheeler1989,Makela2019,Davidson2019}, or effectively imposed through Lagrange multipliers in state-counting frameworks such as loop quantum gravity \cite{Kaul2000,Meissner2004}, the introduction of a chemical potential $\mu$ provides a genuine thermodynamic bias that dynamically selects the occupation fraction, promoting the maximal-entropy configuration to a particular equilibrium point rather than an assumption, as the following section reveals.

\section{Thermodynamic framework as a justification of the partially occupied regime}
\label{sec:thermo}
The entropy derivation above constitutes the central result of this work. It shows that, once the occupied-site sector enters a partially filled regime with finite $n$, the functional dependence of the entropy is no longer an assumption but an unavoidable consequence of the underlying combinatorics. In particular, the emergence of an area law together with logarithmic and inverse-area corrections follows purely from counting arguments, independently of any prior thermodynamic input. This raises a natural question: under what physical conditions does the horizon actually realize such a partially occupied regime? In other words, can the existence of a finite filling fraction $n_0\in(0,1)$ be understood as the outcome of an underlying thermodynamic principle rather than an ad hoc assumption? We now address this point by showing that such a regime admits a natural thermodynamic interpretation. Specifically, we extend the framework of black-hole chemistry by promoting the number of occupied sites $N$ to a thermodynamic variable. Within this enlarged phase space, the partially filled configuration emerges dynamically as an equilibrium state, thereby providing a physical justification for the combinatorial construction introduced above.

\subsection{Black-hole chemistry with a compositional variable}
The framework of black-hole chemistry can be understood as a systematic extension of standard black-hole thermodynamics, motivated by the desire to restore a closer analogy with ordinary thermodynamic systems \cite{Kubiznak_2017}.  In the usual formulation, a static and neutral black hole obey a first law of the form 
\begin{equation}
d M = T\,d S ,
\end{equation}
where $M$ is interpreted as the internal energy of the system. The key motivation for extending this framework is therefore to identify a physically meaningful notion of pressure and its conjugate variable. The crucial observation is that a negative cosmological constant $\Lambda$ naturally behaves as a vacuum energy density, and hence can be associated with a thermodynamic pressure according to
\begin{equation}
P = -\frac{\Lambda}{8\pi}.
\end{equation}
Once this identification is made, the thermodynamic phase space is enlarged by allowing $\Lambda$ to vary. In this extended phase space, consistency with the first law requires the introduction of a conjugate thermodynamic volume,
\begin{equation}
V = \left(\frac{\partial M}{\partial P}\right)_{S,Q,J},
\end{equation}
leading to the generalized first law
\begin{equation}
d M = T\,d S + V dP 
\end{equation}
A central conceptual consequence of this extension is that the ADM mass $M$ is no longer interpreted as internal energy, but rather as enthalpy,
\begin{equation}
M = E + P V,
\end{equation}
representing the total energy required to form the black hole in a background with nonzero vacuum pressure.

The extended framework is further supported by the corresponding Smarr relation, which in four dimensions takes the form
\begin{equation}
M = 2 T S - 2 P V. 
\end{equation}

The Smart law acts as an "integrated" version of the first law of black hole thermodynamics.
Beyond formal consistency, this extension reveals a rich thermodynamic structure. In particular, asymptotically AdS black holes exhibit an equation of state analogous to that of a Van der Waals fluid \cite{Kubiznak_2017},
\begin{equation}
P = \frac{T}{v} - \frac{1}{2\pi v^2},
\end{equation}
with $v$ playing the role of a specific volume. This leads to phase transitions and critical phenomena closely resembling those of ordinary fluids.

From this perspective, black-hole chemistry takes seriously the possibility of enlarging the thermodynamic phase space by promoting quantities that were previously regarded as fixed geometric parameters to dynamical thermodynamic variables. 

In the present work, motivated by the microscopic picture introduced in the previous section, we follow the same philosophy by introducing an additional compositional variable $N$, associated with the number of occupied horizon sites. In direct analogy with the role of $\Lambda$, the variable $N$ is promoted to a thermodynamic degree of freedom, so that the macroscopic state is described by
\begin{equation}
M = M(S,P,N),
\end{equation}
and the corresponding first law is expected to take the form
\begin{equation}
dM = T\,dS + V\,dP + \mu\,dN,
\end{equation}
where $\mu$ is the quantity conjugate to $N$. 
Under geometric scaling $L\to \lambda L$, the standard thermodynamic variables scale as
\begin{equation}
S\to \lambda^2 S,
\qquad
P\to \lambda^{-2}P,
\end{equation}
while the compositional variable scales extensively with the horizon area,
\begin{equation}
N\to \lambda^2 N.
\end{equation}
Since $M\sim L$ in geometrized units \footnote{We work in geometrized units $G=c=\hbar=k_B=1$}, Euler homogeneity yields the Smarr relation
\begin{equation}\label{smarr}
M=2TS-2PV+2\mu N.
\end{equation}

\subsection{Smarr-consistent excitation enthalpy around a finite-filling reference configuration}

The last ingredient of the extended thermodynamic framework is an explicit form, or at least a controlled approximation, for the enthalpy
\begin{equation}
M=M(S,P,N).
\end{equation}
 Rather than postulating an ansatz for $M$, it is therefore natural to regard the Smarr relation as a differential constraint that determines its allowed functional form. More precisely, Eq. \eqref{smarr} can be expressed as
\begin{equation}
M=2S\frac{\partial M}{\partial S}-2P\frac{\partial M}{\partial P}+2N\frac{\partial M}{\partial N},
\end{equation}
which general solution is (see Appendix A)
\begin{equation}
M(S,P,N)=\sqrt{S}\,f(u,v),
\ \
u\equiv PS, \ v\equiv \frac{N}{S},
\end{equation}
where $f(u,v)$ is an arbitrary function of the two scale-invariant combinations $u$ and $v$.  This result has a clear interpretation: the full thermodynamic information is encoded in a function of dimensionless variables, while the overall $\sqrt{S}$ factor is fixed by scaling. In particular, the ratio $v=N/S$ emerges as the natural thermodynamic measure of horizon occupancy, consistently with the filling fraction introduced at the microscopic level.

The entropy analysis of Sec.~\ref{sec:combi} singles out a reference partially occupied configuration
\begin{equation}
N_0(A)=n_0\,\mathcal{N}(A)=n_0\frac{A}{a_p},
\qquad 0<n_0<1,
\end{equation}
with $n_0=\tfrac12$ for the maximal-entropy choice. This configuration plays the role of a thermodynamic background around which fluctuations in the occupation number should be described. To keep the thermodynamic description aligned with this combinatorial saddle, we expand the Smarr-compatible enthalpy around the corresponding reference value
\begin{equation}
v_0\equiv \frac{N_0}{S}=\frac{4n_0}{a_p}.
\end{equation}
Define
\begin{equation}
\Delta N\equiv N-N_0,
\qquad
\Delta v\equiv v-v_0=\frac{\Delta N}{S}.
\end{equation}

At this point, an important consistency requirement enters. In the absence of occupation fluctuations, namely when $N$ is fixed at its reference value, the thermodynamic description must reduce to the standard black-hole chemistry without the compositional sector. This implies that the enthalpy must admit a decomposition into a reference contribution and occupation-dependent corrections. Retaining only the leading occupation-dependent term in the expansion of $f(u,v)$ around $v_0$ gives
\begin{equation}
f(u,v)=f_0(u)+f_1(u)\,\Delta v+O(\Delta v^2),
\end{equation}
so that
\begin{equation}
M(S,P,N)=M_{\rm ref}(S,P)+\delta M(S,P,N),
\end{equation}
with
\begin{equation}
\delta M(S,P,N)=\frac{f_1(u)}{\sqrt{S}}\,(N-N_0)+O\!\left((N-N_0)^2\right).
\end{equation}
Equivalently, this can be written as
\begin{equation}
\delta M(A,P,N)=\frac{\Gamma(u)}{\sqrt{S}}\,(N-N_0)+O\!\left((N-N_0)^2\right),
\end{equation}
where $u\equiv PS$ and
\begin{equation}
\Gamma(u)\equiv f_1(u).
\end{equation}

This structure admits a simple physical interpretation. Near the partially occupied reference configuration, the thermodynamic dressing assigns a finite excitation cost per occupied horizon site,
\begin{equation}
\varepsilon(u)\equiv \frac{\Gamma(u)}{\sqrt{S}},
\end{equation}
so that the leading change in the enthalpy is linear in the deviation $N-N_0$ from the reference occupation. In this sense, the compositional variable $N$ behaves as a genuine extensive degree of freedom, whose fluctuations are governed by an effective single-site excitation energy.

\subsection{Grand-canonical description and occupation fluctuations}

The construction above provides a consistent thermodynamic description of the enthalpy, including the effect of occupation through a linear excitation energy around a partially filled reference configuration. However, this description remains at the level of equilibrium thermodynamics. To complete the picture, it is necessary to incorporate a statistical description of the occupation number $N$. Since $N$ is a genuine thermodynamic degree of freedom with conjugate chemical potential $\mu$, the natural framework is a grand-canonical ensemble, where the horizon area $A$ plays the role of an effective container and the occupation number is allowed to adjust under the control of $\mu$. In this setting, both the equilibrium filling fraction and its fluctuations follow from a single thermodynamic potential.

Up to an irrelevant constant shift proportional to $N_0$, the microstructure grand potential, $\Phi(A,P,T,\mu;N)$ may be written as
\begin{equation}
\Phi=\delta M(A,P,N)-T\,S_{\rm micro}(A,N)-\mu(N-N_0).
\end{equation}
Equilibrium is determined by the stationarity condition
\begin{equation}
\left(\frac{\partial \Phi}{\partial N}\right)_{A,P,T,\mu}=0.
\end{equation}
Using
\begin{equation}
\left(\frac{\partial S_{\rm micro}}{\partial N}\right)_A
=\ln\!\left(\frac{1-n}{n}\right),
\qquad
n=\frac{Na_p}{A},
\end{equation}
we obtain the exact equilibrium relation
\begin{equation}
\mu=\frac{\Gamma(u)}{\sqrt{S}}-T\ln\!\left(\frac{1-n}{n}\right).
\end{equation}
Equivalently,
\begin{equation}
\ln\!\left(\frac{1-n_\star}{n_\star}\right)=\frac{\Gamma(u)/\sqrt{S}-\mu}{T},
\end{equation}
so the equilibrium filling fraction is
\begin{equation}
n_\star=\frac{1}{1+\exp\!\left[(\varepsilon(u)-\mu)/T\right]}.
\end{equation}

This result has a transparent interpretation: the horizon occupation follows a standard binary distribution with a single-site excitation cost $\varepsilon(u)=\Gamma(u)/\sqrt{S}$ and chemical potential $\mu$. In particular, the partially occupied regime
\begin{equation}
0<n_\star<1
\end{equation}
emerges dynamically whenever the effective bias
\begin{equation}
\mu_{\rm eff}\equiv \mu-\frac{\Gamma(u)}{\sqrt{S}}
\end{equation}
remains finite. The reference filling $n_0$ identified from the combinatorial analysis is therefore not imposed, but realized as an equilibrium configuration for a suitable choice of $\mu$.

To characterize the stability of this configuration, we analyze the behavior of the grand potential around the reference occupation. Defining
\begin{equation}
N_0=n_0\,\mathcal{N},
\qquad
\mathcal{N}=\frac{A}{a_p},
\qquad
\Delta N \equiv N-N_0,
\end{equation}
and using the linear excitation enthalpy,
\begin{equation}
\delta M(A,P,N)=\varepsilon(u)\,(N-N_0),
\qquad
u=PS,
\end{equation}
we expand the grand potential around $N_0$. Expanding the combinatorial entropy gives
\begin{eqnarray}
S_{\rm micro}(A,N)
=
S_{\rm micro}(A,N_0)
+
\left.\frac{\partial S_{\rm micro}}{\partial N}\right|_{N_0}\Delta N
\nonumber\\
+\frac{1}{2}
\left.\frac{\partial^2 S_{\rm micro}}{\partial N^2}\right|_{N_0}
(\Delta N)^2
+\cdots .
\end{eqnarray}

If $N_0$ corresponds to the equilibrium configuration, the linear term vanishes, yielding
\begin{equation}
\mu
=
\varepsilon(u)
-
T\ln\!\left(\frac{1-n_0}{n_0}\right).
\end{equation}
At half filling, $n_0=\tfrac12$, this reduces to
\begin{equation}
\mu=\varepsilon(u).
\end{equation}

The quadratic term governs the fluctuations. Since the excitation enthalpy is linear in $N$, the second derivative is purely entropic,
\begin{equation}
\Phi''(N_0)
=
T\left(
\frac{1}{N_0}
+
\frac{1}{\mathcal{N}-N_0}
\right).
\end{equation}
The grand potential therefore takes the Gaussian form
\begin{equation}
\Phi(N)\simeq \Phi(N_0)+\frac{1}{2}\,\Phi''(N_0)\,(N-N_0)^2,
\end{equation}
leading to a Gaussian distribution of occupation-number fluctuations,
\begin{equation}
e^{-\Phi(N)/T}
\propto
\exp\!\left[
-\frac{\Phi''(N_0)}{2T}(N-N_0)^2
\right].
\end{equation}

The variance is
\begin{equation}
\mathrm{Var}(N)=\frac{T}{\Phi''(N_0)}
=
\frac{1}{\dfrac{1}{N_0}+\dfrac{1}{\mathcal{N}-N_0}}
=
\mathcal{N}\,n_0(1-n_0).
\end{equation}

Thus, the equilibrium occupation is fixed by the balance between excitation energy and chemical potential, while the width of the fluctuations is entirely controlled by the combinatorial entropy. This provides a consistent bridge between the microscopic counting and the macroscopic thermodynamic description.

\section{Final comments and conclusions}
\label{sec:summ}

We now summarize the construction and clarify the role played by the thermodynamic sector. The central result of the model is combinatorial: a horizon of area $A$ is assumed to contain $\mathcal{N}(A)=A/a_p$ microscopic sites, of which $N$ are occupied. Once the occupied-site sector realizes a partially occupied regime with finite filling, the binomial counting of configurations yields an entropy of the form
\begin{equation}
S(A)=\frac{A}{4}-\frac{1}{2}\ln A+\sum\limits_{\alpha\in {\rm odd}}\frac{D_\alpha}{A^{\alpha}}+\hbox{constant},
\end{equation}
where the Bekenstein--Hawking term emerges without being imposed as an input, and the logarithmic correction arises from subleading terms in the Stirling expansion. The maximal-entropy configuration fixes the elementary area scale to $a_p=4\ln 2$.

A simple generalization of the previous result to have the term $-\frac{k}{2}\ln A$ coming from the purely \emph{combinatorial} saddle, is to include  $k$
independent occupation sectors. Thus, there would be $k$ equivalent Gaussian prefactors inside Stirling's formula, yielding $k$ times the $-\frac{1}{2}\ln A$ term. This generalization keeps the model purely entropic and does not modify the dominant $A/4$ term.

The thermodynamic framework introduced in this work does not generate the entropy formula itself. Its role is instead to provide a consistent dynamical setting in which the partially occupied regime becomes physically admissible. By extending black-hole chemistry to include a compositional variable $N$, we obtain an enlarged first law with chemical potential $\mu$ and a Smarr-consistent enthalpy $M(S,P,N)$. Expanding around a reference configuration $N_0=n_0A/a_p$, the leading correction to the enthalpy is linear in $N-N_0$ and defines a finite single-site excitation cost.

This structure allows for a statistical description of the horizon occupation. In a grand-canonical ensemble, the equilibrium filling fraction follows a binary distribution controlled by the competition between excitation energy and chemical potential, while fluctuations around equilibrium are Gaussian and entirely determined by the combinatorial entropy. In this sense, the thermodynamic sector provides a dressing of the horizon microstructure, selecting the equilibrium occupation and governing its response to the external variables $(P,T,\mu)$.

A consistency condition arises when considering the large-area limit at fixed pressure. Since $x=PS\propto A$, the excitation cost per occupied site,
\begin{equation}
\varepsilon(u)=\frac{\Gamma(u)}{\sqrt{S}},
\end{equation}
remains finite only if
\begin{equation}
\Gamma(u)\sim \Gamma_\infty\,u^{1/2},
\qquad u\to\infty.
\end{equation}
With this scaling,
\begin{equation}
\frac{\Gamma(u)}{\sqrt{S}}\to \Gamma_\infty\sqrt{P},
\end{equation}
and the equilibrium filling approaches a finite value,
\begin{equation}
n_\star\to \frac{1}{1+\exp\!\left[(\Gamma_\infty\sqrt{P}-\mu)/T\right]}.
\end{equation}
This shows that the thermodynamic dressing does not modify the entropy derived from combinatorics, but ensures that the partially occupied regime persists as the black hole grows.

The model may therefore be interpreted as follows. The leading entropy is governed by the combinatorics of horizon microstates in a partially occupied configuration, while the thermodynamic sector determines which filling is realized in equilibrium and assigns an energetic cost to deviations from it. In this way, the horizon microstructure provides the dominant contribution to the entropy, and thermodynamics supplies the mechanism that renders this structure dynamically consistent within the framework of black-hole chemistry.
\begin{acknowledgments}
 P. B. and E. C. acknowledge financial support from the Generalitat Valenciana through PROMETEO PROJECT CIPROM/2022/13. E. C. is funded by the Beatriz Galindo 
 contract BG23/00163 (Spain). E. M. acknowledges financial support from the POLI Grant 41981.
\end{acknowledgments}

\bibliography{references}

\appendix
\section{Characteristic solution of the Smarr partial differential equation}

For completeness, we derive the general Smarr-compatible form of the enthalpy quoted in the main text.
Starting from
\begin{equation}
M=2S\left(\frac{\partial M}{\partial S}\right)_{P,N}-2P\left(\frac{\partial M}{\partial P}\right)_{S,N}+2N\left(\frac{\partial M}{\partial N}\right)_{S,P},
\end{equation}
we obtain the first-order linear PDE
\begin{equation}
2S M_S-2P M_P+2N M_N-M=0.
\end{equation}
The characteristic equations are
\begin{eqnarray}
&&\frac{dS}{d\lambda}=2S,
\\
&&\frac{dP}{d\lambda}=-2P,
\\
&&\frac{dN}{d\lambda}=2N,
\\
&&\frac{dM}{d\lambda}=M.
\end{eqnarray}
They integrate immediately to
\begin{eqnarray}
&&S(\lambda)=S_0e^{2\lambda},
\\
&&P(\lambda)=P_0e^{-2\lambda},
\\
&&N(\lambda)=N_0e^{2\lambda},
\\
&&M(\lambda)=M_0e^{\lambda}.
\end{eqnarray}
From these solutions we identify the invariants along the characteristics,
\begin{equation}
u\equiv PS,
\qquad
v\equiv \frac{N}{S},
\qquad
w\equiv \frac{M}{\sqrt{S}}.
\end{equation}
The general solution must therefore express $w$ as an arbitrary differentiable function of the two independent invariants $u$ and $v$:
\begin{equation}
w=f(u,v).
\end{equation}
Returning to the original variables, we obtain the most general Smarr-compatible form
\begin{equation}
M(S,P,N)=\sqrt{S}\,f\!\left(PS,\frac{N}{S}\right).
\end{equation}

\end{document}